\documentclass[aps,prd, superscriptaddress,preprintnumbers,nofootinbib]{revtex4}
\usepackage{amsmath,amsfonts,amssymb,amsthm,amstext,amscd,graphicx}
\usepackage[all]{xy}
\usepackage[colorlinks=true]{hyperref}

\def\be{\begin{equation}}
\def\ee{\end{equation}}
\def\arr{\begin{array}{rll}}
\def\ea{\end{array}}
\def\bea{\begin{eqnarray}}
\def\eea{\end{eqnarray}}

\def\N2{$N{=}2$}
\def\diff{{\rm d}}

\newcommand{\casimir}{\widetilde{\mathcal{I}}}
\newcommand{\nnr}{\nonumber \\}

\def\l3{\ell_3}

\begin{document}
\preprint{IPM/P-2018/004}
\vskip 1 cm
\title{{\Large{On integrability of geodesics  in near-horizon extremal geometries:}}\\ Case of Myers-Perry black holes in arbitrary dimensions}
\author{Hovhannes Demirchian}
\email{demhov@bao.sci.am}
\affiliation{ Ambartsumian Byurakan Astrophysical Observatory, Byurakan, 0213,  Armenia}
\author{Armen Nersessian}
\email{arnerses@yerphi.am}
\affiliation{Yerevan Physics Institute, 2 Alikhanian Brothers St., Yerevan  0036 Armenia}
\affiliation{Bogoliubov Laboratory of Theoretical Physics, Joint Institute for Nuclear Research, 141980, Dubna, Russia}
\author{Saeedeh Sadeghian}
\email{ssadeghian@ipm.ir }
\affiliation{ Institute for Research in Fundamental Sciences (IPM), P.O.Box 19395-5531, Tehran, Iran   }
\author{M.M. Sheikh-Jabbari}
\email{jabbari@theory.ipm.ac.ir }
\affiliation{ Institute for Research in Fundamental Sciences (IPM), P.O.Box 19395-5531, Tehran, Iran   }

\begin{abstract} We investigate  dynamics of probe particles moving in the near-horizon limit of extremal Myers-Perry black holes in arbitrary  dimensions. Employing ellipsoidal coordinates we show that this problem is integrable and separable, extending the results of the odd dimensional case discussed in  \cite{non-equal-general}.  We find the general solution of the Hamilton-Jacobi equations for these systems and  present explicit expressions for the Liouville integrals, discuss  Killing tensors and the associated  constants of motion. We analyze special cases of the background near-horizon geometry were the system possesses more constants of motion and is hence superintegrable.  Finally, we consider near-horizon extremal vanishing horizon case which happens for Myers-Perry black holes in odd dimensions and show that geodesic equations on this geometry are also separable and work out its integrals of motion.

\end{abstract}
\maketitle
%

\setcounter{footnote}0
\section{Introduction}
Any dynamical system, particle or field dynamics alike, is classically described by equations of motion and some boundary conditions for the field theory case. The main task in analyzing the system is to solve the equations of motion, which are generically (partial) second order differential equations, and solving them  is generically a formidable task. Symmetries, Noether theorem and constants of motion, are the usual tools facilitating tackling the problem.   In this work we will focus on particle dynamics on certain $d$ dimensional curved backgrounds and question of field theories on such backgrounds are deferred to an upcoming publication.

In a dynamical system with $N$ degrees of freedom and hence a $2N$ dimensional phase space, if  number of  independent symmetries is equal to $N$,  the system is called \emph{integrable} and is usually solvable. If the system possesses  $N+p,$ $1\leq p\leq N-1$, independent symmetries (and hence functionally independent constants of motion),  it is called \emph{superintegrable} and the region it can probe in its $2N$ dimensional phase space is a compact $N-p$  dimensional surface; e.g. see \cite{Kupershmidt:1990ws, Miller:2013gxa, oxana}.

For the question of particle dynamics on a general curved (usually a black hole) background in $d$ dimensions, we are dealing with a $2d$ dimensional phase space. It is an established fact that isometries of the background, the Killing vectors, provide a set of constants of motion. Moreover, reparametrization invariance of the particle action implies that there is always a second rank Killing-tensor whose conserved charge is the mass of particle. For backgrounds of interest, e.g. black holes or their near horizon geometries, usually the number of Killing vectors plus one is less than $d$ and one may wonder if the system is integrable.

The question of integrability of particle dynamics on black hole or near horizon geometries have been extensively analyzed in the literature e.g. see \cite{Carter, conformal-mechanics-BH-1,conformal-mechanics-BH-2,GNS-1, Frolov:2003en, Benenti and Francaviglia, Kubiznak:2007kh,Krtous:2006qy,Frolov:2006pe,Cariglia:2011qb,Hidden-symmetry-NHEK, Lunin}. In particular, it has been shown that the problem is (super)integrable for a large class of black holes. The integrability is often associated with the existence of higher rank, usually second rank, Killing tensor fields \cite{Carter} (see \cite{Frolov:2017kze} for review).

Given an extremal black hole there are general theorems stating that in the near horizon limit we obtain a usually smooth geometry with larger isometry group than the original extremal black hole \cite{NHEG-general}. It is hence an interesting question to explore if this symmetry enhancement yields further independent constants of motion and how it affects the (super)integrability of particle dynamics.
This question, besides the academic interests, is also relevant to some of the observations related to black holes: It is now a well-accepted fact that there are fast rotating black holes in the sky which are well modeled by an extreme Kerr geometry \cite{Extreme-Kerr} and the matter moving around these black holes in their accretion disks are essentially probing the near horizon geometry \cite{Falcke:1999pj}.

{The isometry group of generic stationary extremal black holes in the near horizon  region is shown to have an $SO(2,1)=SL(2,\mathbb{R})$ part \cite{NHEG-general,NHEG-2}. Therefore, particle dynamics on the near horizon extreme geometries possesses  dynamical $0+1$ dimensional conformal symmetry, i.e.  it defines a ``conformal mechanics" \cite{conformal-mechanics-BH-1,conformal-mechanics-BH-2,GNS-1}. This allows to reduce the problem to the study of  system depending on
latitudinal and azimuthal coordinates and their conjugate momenta with the effective Hamiltonian being Casimir of conformal algebra. Such associated systems have been investigated from  various viewpoints  in Refs. \cite{Armen-Tigran} where they were called ``angular (or spherical) mechanics''.

In this work, we continue our analysis of \cite{non-equal-general, Demirchian:2017uvo} and extend the analysis there to Near Horizon Extremal Myers-Perry \cite{mp} (NHEMP) black holes \cite{NHEG-2} in general odd and even dimensions. We discuss the separability of variables, constants of motion  for ``angular mechanics" associated with
these systems and how they are related to the second rank Killing tensors of the background. } While the system is in general integrable, as we show, there are special cases where the system is superintegrable. Moreover, we discuss another interesting case, the Extremal Vanishing Horizon (EVH)  \cite{NHEVH-1} Myer-Perry black holes \cite{NHEVH-MP} and  show the integrability of geodesics in the Near Horizon EVH Myers-Perry (NHEVH-MP) geometries.

The rest of this paper is organized as follows. In section \ref{NHEMP-background-Sec} we present the geometry of near-horizon extremal Myers-Perry black holes in generic even and odd dimensions, and  construct the ``angular mechanics" describing  probe particle dynamics. In this section we set our notations and conventions. In section \ref{Particle-dynamics-Sec} we analyze generic causal curve, massive or massless geodesic, in the NHEMP background. We show that this Hamiltonian system is separable in ellipsoidal coordinate system, work out the constants of motion and establish that the system is integrable. Moreover, we show how the Killing vectors and second rank Killing tensors are related to these constants of motion. In section \ref{Special-cases-Sec} we analyze special cases where some of the rotation parameters of the background NHEMP are equal. In these cases we have some extra Killing vectors and tensors and the system is superintegrable. Section \ref{EVH-Sec} contains the analysis of particle dynamics on the special class of Extremal Vanishing Horizon (EVH) Myers-Perry black holes. We end this note with discussions and further  comments.

\section{NHEMP in arbitrary dimensions; unified description}\label{NHEMP-background-Sec}

The NHEMP metric in both odd and even dimensions in the  Gaussian null coordinates was presented in \cite{NHEG-2}. The NHEMP is a (generically) a smooth solution to vacuum Einstein equations in odd $d=(2N+1)$- and $d=(2N+2)$-dimensions, in general it is specified by $N$ number of rotation parameters $a_i$ (or $N$ angular momenta $J_i$) and has $SL(2,\mathbb{R})\times U(1)^N$ isometry. In Boyer-Lindquist coordinates NHEMP metric has the form
\be\label{NHEMP-metric-rt-coord}
	ds^2=\frac{F_H}{b}\left(-r^2d\tau^2+\frac{dr^2}{r^2}\right)+\sum_{I=1}^{N_\sigma}(r_H^2+a_I^2)d\mu_I^2+\gamma_{ij}D\varphi^i D\varphi^j,
	\qquad
	D\varphi^i\equiv d\varphi^i+\frac{B^i}{b}rd\tau,
\ee
where 
 $N_\sigma=[\frac{d}{2}]=N+\sigma$, i.e. $\sigma=0$ for the odd and $\sigma=1$ for the even dimensions cases,
 $r_H$ is a black hole radius
 which satisfy the equation
 \be
\sum_{I=1}^{N_\sigma}\frac{r^2_H}{r^2_H+a^2_I}=\frac{1+2\sigma}{1+\sigma}, \qquad{\rm with}\quad a_{N+1}=0
\label{ai}\ee
  and,\footnote{There seems to be a minor typo in the exressions for NHEMP metrics given in \cite{NHEG-2}, which we have corrected here.}
\begin{gather}
	F_H=1-\sum_{i=1}^{N}\frac{a_i^2\mu_i^2}{r_H^2+a_i^2},
	\qquad b=\frac{1}{r_H^2}\left(\sum_{i=1}^{N}\frac{\sigma\, r_H^2}{r_H^2+a_i^2}+4\sum_{i<j}^{N}\frac{r_H^2}{r_H^2+a_i^2}\frac{r_H^2}{r_H^2+a_j^2}\right),\qquad	B^i=\frac{2r_Ha_i}{(r_H^2+a_i^2)^2},\\
	\gamma_{ij}=(r_H^2+a_i^2)\mu_i^2\delta_{ij}+\frac{1}{F_H}a_i\mu_i^2a_j\mu_j^2,
	\qquad \sum_{I=1}^{N_\sigma}\mu_I^2=1.
\end{gather}
In our notations  lowercase Latin indices $i, j$ which run from $1$ to $N$ and uppercase Latin indices $I,J$ which run over
1 to  $N_\sigma$ and $r_H$ satisfies
%

For the case when all $a_i$ take generic non-zero values\footnote{The case when one of the $a_i$ is zero is the EVH case we will discuss separately in section \ref{EVH-Sec}.} it is convenient to introduce
new parameters $m_i$
\be
m_i=\frac{r_H^2+a_i^2}{r_H^2} > 1,\qquad m_{N+1}=1\;\quad  \text{and} \quad \sum^{N_\sigma}_{I=1}\frac{1}{m_I}= \frac{1+2\sigma}{1+\sigma},
\ee
  and re-scaled coordinates $x_I$,
\be
 x_I=\sqrt{m_I}\mu_I \;:\quad \sum_{I=1}^{N_\sigma}\frac{x^2_I}{m_I}=1.
\ee
In these terms the near-horizon metrics reads
\be\label{m2}
	\frac{ds^2}{r^2_H}=A(x)\left(-r^2d\tau^2+\frac{dr^2}{r^2}\right)+\sum_{I=1}^{N_\sigma} dx_Idx_I+
	\sum_{i,j=1}^N\tilde{\gamma}_{ij}x_i x_j D\varphi^iD\varphi^j, \qquad   D\varphi^i\equiv d\varphi^i+k^ird\tau,
\ee
where
\be\label{gamma}
\begin{gathered}
	A(x) ={\frac{\sum_{I=1}^{N_\sigma} x^2_I/m^2_I}{\frac{\sigma}{1+\sigma}+4\sum_{i<j}^{N}\frac{1}{m_i}\frac{1}{m_j}}}, \quad
	\tilde{\gamma}_{ij}=\delta_{ij}+ \frac{1}{\sum_I^{N_\sigma} x_I^2/m^2_I}\frac{\sqrt{m_i-1}x_i}{m_i}  \frac{\sqrt{m_j-1}x_j}{m_j},\\
k^i=\frac{2\sqrt{m_i-1}}{m^2_i(\frac{\sigma}{1+\sigma}+4\sum_{k<l}^{N}\frac{1}{m_k}\frac{1}{m_l})},
\end{gathered}
\ee
with
\be\label{eq:restr_on_x}
\sum_{I=1}^{N_\sigma} \frac{x^2_I}{m_I}=1, \qquad \sum_{I=1}^{N_\sigma} \frac{1}{m_I}=\frac{1+2\sigma}{1+\sigma}.
\ee
With this unified description at hands we are ready to describe probe particle dynamics.

\subsection{Probe-particle dynamics}

The metric \eqref{m2} has $SL(2,\mathbb{R})$ isometry group and hence  the particle dynamics on this background  exhibits  dynamical conformal symmetry; we are dealing with a ``conformal mechanics" problem \cite{conformal-mechanics-BH-1,conformal-mechanics-BH-2, GNS-1}. Let us denote the three generators of this $sl(2,\mathbb{R})$ algebra by $H,D,K$, and its Casimir by ${\cal I}$:
\be
\{H,D\}=H, \quad \{H,K \}=2D, \quad \{D,K \} =K,\qquad \mathcal{I}=HK-D^2.
\label{confalg}\ee

The mass-shell equation for a particle of mass $m_0$ moving in the background metric
 \be\label{mass-shell-1}
 m^2_0=-\sum_{A,B=1}^{2N+1+\sigma}g^{AB}p_A p_B,
 \ee
leads to the following expression
\be
\label{eq:mass_shell} m_0^2r_H^2=\frac{1}{A}\left[\left(\frac{p_0}{r}-\sum_{i=1}^{N}k_ip_{\varphi_i}\right)^2-(rp_r)^2\right]-\sum_{a,b=1}^{N_\sigma-1}h^{ab}p_ap_b
-\sum_{i,j=1}^{N}\tilde{\gamma}^{ij}\frac{p_{\varphi_i}}{x_i}\frac{p_{\varphi_j}}{x_j},
\ee
where
\be
h^{ab}=\delta^{ab}-\frac{1}{\sum\limits_{I=1}^{N_\sigma} x^2_I/m^2_I}\frac{x_a}{m_a}\frac{x_b}{m_b},\qquad
\tilde{\gamma}^{ij}=\delta^{ij}-x_i\frac{\sqrt{m_i-1}}{m_i}x_j\frac{\sqrt{m_j-1}}{m_j},\qquad a,b=1,\cdots N_\sigma-1,\ i,j=1,\cdots, N.
\ee
Using \eqref{eq:mass_shell}, as in \cite{non-equal-general}, we can construct the Hamiltonian $H=p_0$ and the other generators of the conformal algebra
\begin{gather}
\label{eq:Hamiltonian_initial}
H= r\left(\sqrt{ L(x_a, p_a, p_{\varphi_i}) +(rp_r)^2}+\sum_{i=1}^N k_ip_{\varphi_i}\right),\\
D=r p_r,\qquad
K=\frac{1}{r} \left(\sqrt{ L(x_a, p_a, p_{\varphi_i}) +(rp_r)^2}- \sum_{i=1}^N k_ip_{\varphi_i}
\right),
\end{gather}
where
$$
 L(x_a, p_a, p_{\varphi_i})= A\left(m_0 r_H^2+ \sum_{a,b=1}^{N_\sigma-1}h^{ab}p_ap_b
+\sum_{i,j=1}^{N}\tilde{\gamma}^{ij}\frac{p_{\varphi_i}}{x_i}\frac{p_{\varphi_j}}{x_j} \right),
$$
and the momenta $p_a, p_{\varphi_i}, p_r$ are conjugate to $x_a, \varphi_i, r$ with the canonical Poisson brackets
\be
\label{Poisson-bracket}
\{p_a,x_b\}=\delta_{ab}, \qquad \{p_{\varphi_i},\varphi_j\}=\delta_{ij},\qquad\{p_r, r\}=1.
\ee
Thus, the   Casimir element of the conformal algebra  reads
\be
\label{L}
	\mathcal{I}=A\left[\sum_{a,b=1}^{N_\sigma-1}h^{ab}p_ap_b+\sum_{i=1}^{N}\frac{p_{\varphi_i}^2}{x_i^2}+g_0\right]- \mathcal{I}_0
\ee
where
\be
\label{consts}
g_0=-\left(\sum_{i=1}^N \frac{\sqrt{m_i-1}p_{\varphi_i}}{m_i}\right)^2+m^2_0r^2_H,
\qquad
\mathcal{I}_0=\left(\sum_i^N k_ip_{\varphi_i}\right)^2.
\ee
In an appropriately chosen frame $H$ can be written in formally nonrelativistic form
\cite{conformal-mechanics-BH-2,GNS-1}
\be\label{nonrel}
H=\frac{p^2_R}{2}+\frac{2\mathcal{I}}{R^2},
\ee
where $R=\sqrt{2K}$, $ p_R=\frac{2D}{\sqrt{2K}}$ are  the effective ``radius" and its canonical conjugate ``radial momentum''.
As we will show below the Casimir  $\mathcal{I}$  encodes all the essential information about the system of particle on these backgrounds.
The Casimir   $\mathcal{I}$ \eqref{L} is at most quadratic in momenta canonically conjugate to the remaining angular variables and it can conveniently be viewed as the Hamiltonian of a reduced ``angular/spherical mechanics" \cite{Armen-Tigran}  describing motion of particle on some curved background.
Note that the  ``time parameter''    conjugate  to $\mathcal{I}$  is different   than the time  parameter $\tau$ appeared in metric \eqref{m2} whose conjugate variable is $H=p_0$. 
See  \cite{SIGMA} for more detailed discussions.

Since  the azimuthal angular variables $\varphi^i$ are cyclic, corresponding conjugate momenta $p_{\varphi_i}$ are constants of motion.
We then remain with a reduced $(N_\sigma-1)$-dimensional system described by Hamiltonian \eqref{L} and $x_a$ variables and their conjugate momenta.

\section{Fully Non-isotropic case}\label{Particle-dynamics-Sec}

To show that the angular/spherical mechanics system is integrable, we show that it is separable in the ellipsoidal coordinates
when  we are dealing with cases where all parameters $m_i$ are non-equal.
The ellipsoidal coordinates $\lambda_I$ for odd and even dimensions are then defined as
\be
\label{xN}
x^2_I=(m_I-\lambda_I)\prod_{J=1, J\neq I}^{N_\sigma}\frac{m_I-\lambda_J}{m_I-m_J},\qquad \lambda_{N_\sigma}  < m_{N_\sigma}  <  \ldots < \lambda_2  < m_2  < \lambda_1 < m_1.
\ee
To resolve the condition $\sum_{I=1}^{N_\sigma} \frac{x^2_I}{m_I}=1$ we choose $\lambda_{N_\sigma}=0$ and hence there are $N_\sigma-1$ independent $\lambda_I$ variables, which will be denoted by $\lambda_a$.

In these coordinates the  angular Hamiltonian $\mathcal{I}$ (shifted by a constant and appropriately rescaled) reads
\be\label{eq:conf_ham_odd}
	\tilde{\mathcal{I}}=\lambda_1\ldots\lambda_{N_\sigma -1}\left[ - \sum_a^{N_\sigma-1}\frac{{4\prod_{I=1}^{N_\sigma}(m_I-\lambda_a) }\pi^2_a}{\lambda_a\prod_{b=1,a\ne b}^{N_\sigma-1}(\lambda_b-\lambda_a)}+\sum_{i=1}^{N_\sigma}\frac{{g}^2_I}{\prod_{a=1}^{N_\sigma-1}(m_I-\lambda_a)} +g_0\right],
\ee
where
\be	
	\tilde{\mathcal{I}}\equiv  \left({\mathcal{I}+\mathcal{I}_0}\right)\left(\frac{\sigma}{1+\sigma}+4\sum_{k<l}^{N}\frac{1}{m_k}\frac{1}{m_l}\right)\prod_{i=1}^{N}m_i,\qquad \mathcal{I}_0=\left(\sum_i^N k_ip_{\varphi_i}\right)^2,
\ee
with
\be
{g}^2_I =\frac{p^2_{\varphi_I}}{m_I}\prod_{J=1,J\neq I}^{N_\sigma} (m_I-m_J),
	\qquad	g_{N+1}=p_{\varphi_{N+1}}\equiv 0,
\ee
and $\{\pi_a,\lambda_b\}=\delta_{ab}$, $\{p_{\varphi_i},\varphi_j\}=\delta_{ij}$.

 The level surface  of angular Hamiltonian \eqref {eq:conf_ham_odd}, $\tilde{\mathcal{I}}={\cal E}$,    can be conveniently represented through
\be\label{HJO}	
\sum_{a=1}^{N_\sigma -1}\frac{R_a- \mathcal{E}}{\lambda_a\prod_{b=1,a\ne b}^{N_\sigma -1}(\lambda_b-\lambda_a)}=0,
\ee
 where\footnote{Note that $R_a\lambda_a\rightarrow R_a$ and $\nu_a \rightarrow F_{a+1}$ replacements have been assumed in the  current paper compared to \cite{non-equal-general}.}
\be
	R_a\equiv -4\prod_{I=1}^{N_\sigma}(m_I-\lambda_a){\pi_a^2}+(-1)^{N_\sigma}\sum_{I=1}^{N_\sigma}\frac{g_I^2\lambda_a}{m_I-\lambda_a}-g_0(-\lambda_a)^{N_\sigma-1},
\label{24}\ee
and we  used the  identities
\be\label{26}
\frac{1}{\prod_{a=1}^{N_\sigma-1}(\lambda_a-\kappa)} =   \sum_{a=1}^{N_\sigma-1}\frac{1}{\prod_{b=1;a\ne b}^{N_\sigma-1}(\lambda_b-\lambda_a)}\frac{1}{\lambda_a-\kappa},\quad
\frac{1}{\lambda_1\ldots\lambda_{N_\sigma-1}}= \sum_{a=1}^{N_\sigma-1} \frac{1}{\prod_{b=1; b\neq a}^{N_\sigma-1}(\lambda_b-\lambda_a)}\frac{1}{\lambda_a}.
\ee
We can rewrite the expression \eqref{HJO} in more useful form, recalling the identities,
\be\label{eq:relation_1}
\sum_{a=1}^{N_\sigma -1}\frac{\lambda_a^{\alpha}}{\prod\limits_{\substack{b=1\\{b\ne a}}}^{N_\sigma -1}(\lambda_a-\lambda_b)}=\delta_{\alpha,N_\sigma-2} \qquad\alpha=0,...,N_\sigma-2.
\ee
Multiplying both sides of \eqref{eq:relation_1} by arbitrary constants $\nu_\alpha$ and  adding  to  \eqref{HJO}, we get
\be\label{HJ1}
\sum_{a=1}^{N_\sigma -1}\frac{R_a(\pi,\lambda)
-\sum_{c=1}^{N_\sigma-1}\nu_{c-1}\lambda^{c-1}_a}{\lambda_a\prod_{b=1,a\ne b}^{N_\sigma -1}(\lambda_b-\lambda_a)}=0,\qquad \nu_0=\mathcal{E}.
\ee

Equipped with the above we can solve the  Hamilton-Jacobi equations
\be
{\mathcal{E}}(\lambda_a,\frac{\partial S_{gen}}{\partial \lambda_a} )=\nu_0,
\ee
and obtain the generating function $S_{gen}$ depending on $N_\sigma-1$ integration constants (i.e. the  general solution of Hamilton-Jacobi equation).
To this end we substitute in \eqref{HJ1}
\be
\pi_a=\frac{\partial S_{gen}}{\partial \lambda_a},
\ee
and  choose the ansatz
\be\label{S-gen}
S_\text{gen}(\lambda_1,\dots,\lambda_{N_\sigma-1})=\sum_{a=1}^{N_\sigma-1} S(\lambda_a).
\ee
This reduces the Hamilton-Jacobi equation to a set of $N_\sigma -1$ ordinary differential equations
\be
\label{31}
R\Big(\lambda_a,\frac{d S(\lambda_a)}{d\lambda_a}\Big)
-\sum_{b =1}^{N_\sigma -1}\nu_{b-1}\lambda^{b-1}_a=0,
\ee
or in an  explicit form,
\be
-{4}\left(\frac{dS(\lambda_a)}{d\lambda_a}\right)^2\prod_{I=1}^{N_\sigma}{(m_I-\lambda_a)}
+ (-1)^{N_\sigma}\sum_{I=1}^{N_\sigma}\frac{{ g}^2_I {\lambda_a}}{m_I-
\lambda_a}-g_0(-\lambda_a)^{N_\sigma-1}-\sum_{b =1}^{N_\sigma -1}\nu_{b-1}\lambda^{b-1}_a=0.
\label{part}\ee
Hence,  the analytic solution to the Hamilton-Jacobi equation is given through the generating function \eqref{S-gen}  with
\be\label{Generating-function}
S(\lambda,\nu_a )=\frac 12\frac{d\lambda}{\sqrt{
\prod_{I=1}^{N_\sigma}{(m_I-\lambda)}}}
\sqrt{
(-1)^{N_\sigma}\left[\sum_{I=1}^{N_\sigma}\frac{{ g}^2_Im_I}{m_I-\lambda}
+g_0\lambda^{N_\sigma-1}-\sum_{i=1}^N{ g}^2_i\right]
-\sum_{b =1}^{N_\sigma -1}\nu_{b-1}\lambda^{b-1}}\ .
\ee
Then, differentiating with respect to constants $\nu_a$,  we can get the explicit solutions of the equations of motion
\be
\tau=\frac{\partial S_{gen}}{\partial \nu_0}\equiv\frac{\partial S_{gen}}{\partial {\cal E}},\qquad  c_a=\frac{\partial S_{gen}}{\partial \nu_a}
\ee

To include the dynamics of azimuthal coordinates $\varphi_i$ we have to consider the generating function
$
S_{tot}=S_{gen}+\sum_{i=1}^N p_{\varphi_i}\varphi_i,
$
where we take into account functional dependence of  $g_0, g_i$ from $p_{\varphi_i}$.
This yields the solutions for azimuthal coordinates
\be
\varphi_i=-\frac{\partial S_{gen}}{\partial p_{\varphi_i}}.
\ee
Thus, we get the  solutions of the angular sector of generic NHEMP with non-equal non-vanishing rotational parameters.

\subsection{Constants of motion }

The expressions for commuting constants of motion $F_a$ can be found from \eqref{31}, by expressing constants $\nu_a$ in terms of  $\lambda_a, \pi_a=\partial S_{gen}/\partial\lambda_a$:
\be
\sum_{b =1}^{N_\sigma-1}F_b\lambda^{b-1}_a=R_a(\pi_a,\lambda_a)\qquad
\Longleftrightarrow \qquad
\begin{pmatrix}
	1 & \lambda_1 & \lambda_1^2 & \cdots & \lambda_1^{N_\sigma-2}
	\\
	1 & \lambda_2 & \lambda_2^2 & \cdots & \lambda_2^{N_\sigma-2}
	\\
	\vdots & \vdots & \vdots & \ddots &\vdots
	\\
	1 & \lambda_{N_\sigma-1} & \lambda_{N_\sigma-1}^2 & \cdots & \lambda_{N_\sigma-1}^{N_\sigma-2}
\end{pmatrix}
\begin{pmatrix}
	F_1
	\\
	F_2
	\\
	\vdots
	\\
	F_{N_\sigma -1}
\end{pmatrix}
=
\begin{pmatrix}
	R_1
	\\
	R_2
	\\
	\vdots
	\\
	R_{N_\sigma -1}
\end{pmatrix},
\ee
where
 $ R_a(\lambda_a, \pi_a )$
are given by \eqref{24}. Integrals of motion  are the solutions to this equation and may  be expressed via the inverse Vandermonde matrix, explicitly,
\be
\label{eq:F_general}
F_\alpha =(-1)^{\alpha -1}\sum_{a=1}^{N_\sigma-1}R_a\ \frac{ A_{N_\sigma -\alpha-1}^{\ne a}}{\prod\limits_{\substack{b=1\\b\ne a}}^{N_\sigma-1}(\lambda_b-\lambda_a)}, \quad \alpha =1,...,N_\sigma -2,\qquad F_{N_\sigma-1}=\sum_{a=1}^{N_\sigma-1}\frac{R_a}{\prod\limits_{\substack{b=1\\b\ne a}}^{N_\sigma-1}(\lambda_a-\lambda_b)},
\ee
where
\be
	A_\alpha^{\ne a}\equiv\sum\limits_{\substack{1\le k_1 < ... < k_{\alpha}\\k_1,...,k_{\alpha}\ne a}}^{N_\sigma-1}\lambda_{k_1}\ ...\ \lambda_{k_{\alpha}}.
\ee

After tedious transformations  on can  rewrite these expressions in $x_a, \varphi_i$  coordinates,
\be
	\label{eq:integrals}
	F_{a}=(-1)^a\sum_{b,c=1}^{N_\sigma-1}K^{bc}_{(a)}(x)p_bp_c- \sum_{i,j=1}^{N}L^{ij}_{(a)}p_{\varphi_i}p_{\varphi_j}+(-1)^{a-1}A_{N_\sigma-a}m_0^2r_H^2,
\ee
where
\bea
	&&K^{bc}_{(a)}=
	\left(
	\sum_{\alpha=0}^{N_\sigma-a-1}(-1)^{N_\sigma+\alpha-a}A_{\alpha}m_b^{N_\sigma-\alpha-a}+
	x^2_b\sum _{\alpha=1}^{N_\sigma-a-1}(-1)^{\alpha}M_{N_\sigma-\alpha-a-1}^{\ne b}m_b^{\alpha }
	\right)\delta^{bc}
	+
	M_{N_\sigma-a-1}^{\ne b,c}x_bx_c\\
	&&L^{ij}_{(a)}=\left((1-\delta_a^1)\sum_{\alpha=1}^{N_\sigma-a}(-1)^{N_\sigma+\alpha}A_{\alpha-1}m_i^{N_\sigma-a-\alpha+1}-\delta_a^1A_{N_\sigma-1}\right)\frac{\delta^{ij}}{x^2_i}
+(-1)^{a-1}A_{N_\sigma-a}\frac{\sqrt{m_i-1}}{m_i}\frac{\sqrt{m_j-1}}{m_j}
\eea
%
with
\be
	\label{eq:f}
	A_a(x_i,m_j)\equiv \sum\limits_{1\le k_1 < ... < k_{a}}^{N_\sigma-1}\lambda_{k_1}\ ... \ \lambda_{k_{a}}=-\sum_{i=1}^{N_\sigma}x^2_iM^{\ne i}_{a-1}+
\sum\limits_{\substack{1\le k_1 < ... < k_{a}}}^{N_\sigma}m_{k_1}\ ... \ m_{k_{a}}
, \qquad a=1,\ldots,N_\sigma-1,
\ee
and
\be
	M^{\ne a_1, ..., a_j}_i\equiv \sum\limits_{\substack{1\le k_1 < ... < k_{i}\\k_1,...,k_i\ne a_1, ..., a_j}}^{N_\sigma}m_{k_1}\ ... \ m_{k_{i}} , \qquad j=0,\ldots,N_\sigma-1,\quad i=1,\ldots,N_\sigma-j.
\ee
It is also assumed that
\be
	A_0\equiv 1, \qquad M^{\ne a_1, ..., a_j}_0\equiv 1.
\ee
One can check that in odd dimensions in the special cases of $F_{N-1},\ F_{N-2}$ and $F_{N-3}$, the above reduce to the corresponding integrals of motion given in \cite{Demirchian:2017uvo}.
One can also check,  that  simply requiring the rotation parameters to be equal in these expressions, one does not recover all the integrals of the special case of $a_i=a, \forall i$  NHEMP. In such special cases all of the first integrals of the spherical mechanics of generic (non-equal $a_i$) case  transform into the Hamiltonian of the spherical mechanics of the equal $a_i$ case. So, to obtaining the Liouville integrals in the isotropic case we need to develop more sophisticated contraction procedure.

 We also note that  the above expressions for the constants of motion  were found  in  the ellipsoidal coordinates introduced  for the special case of {\sl non-equal} rotational parameters $a_i$.
However, we then written  them   in the initial coordinates, they hold   for generic nonzero values of the rotation parameters $a_i$.
We will analyze the special cases where some of the $a_i$ or $m_i$ are equal in section \ref{Special-cases-Sec} and when one of them is vanishing in section \ref{EVH-Sec}.

\subsection{Killing tensors}
In previous subsection  we presented the constants of motion in the form  demonstrating their  explicit dependence on the momenta $p_a, p_{\varphi_i}$.
To represent \eqref{eq:integrals} through the respective $N_\sigma$ second rank Killing tensors, one can replace the last term proportional to $m_0^2$ from  the mass-shell equation \eqref{mass-shell-1},  \eqref{eq:mass_shell}.  Note also that the $F_a,\  a=1,\ldots.N_\sigma-1$, provides $N_\sigma-1$ one constants of motion. We can then add $F_{N_\sigma}$ to this collection, which is proportional to the mass with the corresponding second rank killing tensor being the inverse metric, i.e.
\be	
	F_{N_\sigma}=(-1)^{a-1}\left(r^2_H\sum_{A,B=1}^{2N+1+\sigma}g^{AB}p_A p_B-\left(\sum_{i=1}^N \frac{\sqrt{m_i-1}p_{\varphi_i}}{m_i}\right)^2\right),
\ee
where we assumed	$M_{-1}^{\ne b,c}=0$.

To get the expression for Killing tensors, we should simply replace the momenta by the respective vector fields, $p_A\to \frac{\partial}{\partial x^A}$.
That is,  in the coordinates $(x_a, \varphi_a)$  where the constants of motion  \eqref{eq:integrals} are written,
one should replace
$$
p_a \to \frac{\partial}{\partial x^a}, \quad p_{\varphi_i} \to \frac{\partial}{\partial \varphi^i},\quad p_r\to \frac{\partial}{\partial r},\quad p_0\to\frac{\partial}{\partial \tau}.
$$
In ellipsoidal coordinates the  above presented  $N_\sigma-2$ Killing tensors read
\be
K_{a}=\sum_\alpha A_\alpha^{\neq a}\, h^\alpha\, \left(\partial_{\lambda_\alpha}\right)^2+\sum_I\sum_\alpha \frac{A_\alpha^{\neq a}\,{\prod_{J\neq I}\,} (m_J-m_I)}{m_I(m_I-\lambda_a){\prod_{b}\,}'(\lambda_b-\lambda_a)}\,\left(\partial_{\varphi_I}\right)^2+\frac{A^{a}}{A(\lambda)}\left(-\frac{1}{r^2}\left(\partial_\tau\right)^2+r^2 \left(\partial_r\right)^2\right).
\ee
Thus,  we have $N+1$ mutually commuting Killing vectors $\partial/\partial\varphi_i, \partial /\partial_\tau$
 and $N_\sigma$ Killing tensors, summing up to $d=N_\sigma+N+1$ and hence the system is integrable. One may check that our expressions for the Killing tensors match with those appeared in \cite{Hidden-symmetry-NHEK, Kolar:2017vjl}  after taking the near-horizon limit. We note that the two extra Killing vectors of the $SL(2,\mathbb{R})$ part of the isometry which appear in the near horizon limit and in the coordinates of
 \eqref{m2} take the form,
 \be
r\frac{\partial}{\partial r}-\tau\frac{\partial }{\partial \tau},\qquad (\tau^2+\frac{1}{r^2})\frac{\partial }{\partial \tau}-2\tau r\frac{\partial}{\partial r}-\frac{2}{r}\sum_{i=1}^N \frac{\partial}{\partial\varphi_i},
\ee
do not yield new independent constants of motion.

\section{Isotropic  and partially isotropic cases}\label{Special-cases-Sec}

When some of the $a_i\neq 0$'s are equal the geometry \eqref{NHEMP-metric-rt-coord} exhibits a bigger isometry group than $SL(2,\mathbb{R})\times U(1)^N$;  depending on the number of equal $a_i$'s the $U(1)^N$ part is enhanced to a rank $N$ subgroup of $U(N)$. This larger isometry group brings larger number of Killing vectors and tensors and one hence expects the particle dynamics for these cases to become  a superintegrable system. This is what we will explore in this section and construct the corresponding conserved charges.

\subsection{The fully isotropic, equal $m_i$ case}\label{fully-isotropic-sec}

When all of the rotational parameters coincide, the Hamiltonian of probe particle  reduces to the system on sphere and admits separation of variables in spherical coordinates \cite{GNS-1}.
It can be checked that in this case, the Hamiltonian of the reduced mechanics derived from \eqref{L} transforms into the corresponding mechanics with equal parameters derived in \cite{GNS-1}
for both odd and even dimensional cases.
Notice,  that in this limit the difference between even and odd cases becomes visible:
\begin{itemize}
\item
In the odd case, $\sigma=0$, isotropic limit corresponds to the choice $m_i=N$ , $i=1,\ldots, N$.
As a result, the angular Hamiltonian \eqref{L} which we will denote it by ${\cal I}_N$ takes the form
\be
\mathcal{I}_N=\sum\limits_{a,b=1}^{N-1}(N\delta_{ab}-{x_a x_b})p_{a}p_{b}+N\sum\limits_{i=1}^{N}\frac{p_{\varphi_i}^2}{x_i^2},\qquad \sum_{i=1}^N x_i^2=N.
\label{oddIsoHam}\ee
For the fixed $p_{\varphi_i}$ configuration space of this system is $(N-1)$-dimensional sphere, and   the  Hamiltonian defines specific generalization
of the Higgs oscillator, which is also known as a Rossochatius system \cite{Rosochatius}.

\item In the even case, $\sigma=1$, one has $m_i=2N$ when $i=1,\ldots, N$ and $m_{N+1}=1$, i.e. we can't choose all parameters $m_I$ be equal.
As a result, the angular Hamiltonian \eqref{L} reads
\be
	 \mathcal{I}_N=\sum\limits_{i,j=1}^{N}(\eta^2\delta_{ij}-{x_ix_j})p_{i}p_{j}
	 +\sum\limits_{i=1}^{N}\frac{\eta^2p_{\varphi_i}^2}{x_i^2}
	 +\omega\sum_{i=1}^Nx_i^2,
\label{evenIsoHam}\ee
where
\be
	\eta^2=4N^2- (2N-1)\sum_{i=1}^Nx_i^2,\qquad
	\omega=\left(1-\frac{1}{2N}\right)^2\sum_{i,j=1}^{N}p_{\varphi_i}p_{\varphi_j}-m_0^2\,(2N-1).
\ee
In the case of even dimension configuration space fails to be sphere (even with  fixed $p_{\varphi_i}$).
\end{itemize}
What is important is that both systems admit separation of variables in spherical coordinates.
Namely, by recursively introducing spherical coordinates
\be
	x_{N_\sigma}=\sqrt{N_\sigma}\cos\theta_{N_\sigma-1}, \qquad x_{a}=\sqrt{N_\sigma}{\tilde x}_a\sin\theta_{N_\sigma-1}, \qquad \sum_{a=1}^{N_\sigma-1}{\tilde x}_a^2=1,
\label{sphin}\ee
we get the following recurrent formulae for the constants  of motion
\bea
	\label{eq:iso_integrals_odd}
	&\sigma=0:&\mathcal{I}_{odd}=F_{N-1},\quad
	F_a=p_{\theta_{a}}^2+\frac{p_{\varphi_{a+1}}^2}{\cos^2\theta_{a}}+\frac{F_{a-1}}{\sin^2\theta_{a}},\quad
	F_0=p_{\varphi_1}^2\label{oddconstseq}\\
&\sigma=1 : & \mathcal{I}_{even}=2 N p^2_{\theta_{N}}+\nu\sin^2\theta_{N} + \left(2 N\cot^2\theta_{N}+1\right)F_{N-1},
\label{evenconsteq}
\eea
It is clear, that  $F_1,\ldots, F_{N_\sigma -1}$ define complete set of Liouville   constants of motion and  the $\sigma=1$ system contains $\sigma=0$ as a subsystem.
Moreover,  the  Rosochatius system (angular Hamiltonian for $\sigma=0$ case with fixed $p_{\varphi_i}$) is superintegrable: it has $N-2$ additional functionally independent constants of motion  defined by the expression
\be\label{eq:hidden_integrals}
 I_{a,a-1}=\left(p_{\theta_{a-2}}\sin\theta_{a-2}\cot\theta_{a-1}-p_{\theta_{a-1}}\cos\theta_{a-2}\right)^2+
\left(p_{\varphi_{a-1}}\frac{\cot\theta_{a-1}}{\cos\theta_{a-2}}+p_{\varphi_a}\cos\theta_{a-2}\tan\theta_{a-1}\right)^2.
\ee
When $p_{\varphi_i}$ are not fixed, the system is $(N_\sigma -1+ N)$-dimensional one.  In that case, from its action-angle formulation \cite{GNS-1} one can observe, that
it remains maximally superintegrable for $\sigma=0$, i.e. possesses $4N-3$ constants of motion:
Besides   $2N-3$ constants of motion  given by \eqref{oddconstseq} and \eqref{eq:hidden_integrals}, and
the $N$ commuting integrals  $p_{\varphi_i}$ (associated with axial Killing vectors), there are  $N$ additional  constants of motion with quadratic term mixing $p_{\theta_a}$ and $p_{\varphi_i}$; i.e. $N$ second rank Killing tensors in $\partial_{\theta_a}\partial_{\varphi_i}$ direction.
When $\sigma=1$, the system is $2N$-dimensional, and has  $4N-2$ integrals, i.e., as lacks one integral from being maximal superintegrable.

From these constant of motion one can readily read the associated Killing vectors and second rank Killing tensors.
 Hence, isotropic system has $N+1$ mutually commuting Killing vectors and  $d-3=2N+\sigma-2$ Killing tensors, and an additional $N$ non-commuting second rank Killing tensors.


For more detailed  analysis of the isotropic case see \cite{GNS-1}. Here we present it mainly to set the conventions we use in the study of ``intermediate case", when only some of the rotation parameters are equal to each other.

\subsection{Partially isotropic case in odd dimension}

Let's start with the simpler odd dimensional system, $\sigma=0$, with $p=N-l$  nonequal rotation parameters and  $l$ equal ones:
\be
	m_1\ne m_2\ne\ldots\ne m_p\ne m_{p+1},\quad  m_{p+1}=m_{p+2}= \ldots = m_{N}\equiv\kappa.
\ee
Starting from the metric \eqref{m2} we will construct the Hamiltonian for the reduced mechanics by introducing spherical and ellipsoidal coordinates.
 Spherical coordinates $\{y,\ \theta_i\},\ i=1\ldots l-1$ will be introduced for the $l$ latitudinal coordinates $x_{p+1},\ldots, x_{N}$ corresponding to the equal rotational parameters
\be
\label{eq:shpere_intremediate_special}
\begin{gathered}
	x_{p+1}=y\prod_{i=1}^{l-1}\sin{\theta_{i}},\qquad x_{p+a}=y	\cos{\theta_{a-1}}\prod_{i=a}^{l-1}\sin{\theta_{i}},\quad x_{p+l}=y \cos{\theta_{l-1}},\qquad a=2,\ldots, l-1.
\end{gathered}
\ee
Hence,
\be
	\sum_{a=1}^{l}\frac{x^2_{p+a}}{m_{p+a}}=\frac{y^2}{\kappa}, \qquad
	\sum_{a=1}^{l}(dx_{p+a})^2=(dy)^2+y^2\ d\Omega_{l-1},
\ee
with $d\Omega_{l-1}$ being the metric on $(l-1)$-dimensional sphere: $d \Omega_{l-1}=d\theta_{l-1}^2+\sin^2\theta_{l-1} d \Omega_{l-2}$.

Performing the coordinate transformation \eqref{eq:shpere_intremediate_special} in \eqref{NHEMP-metric-rt-coord}, it is seen that the radial coordinate $y$ of the spherical subsystem behaves very much like the other latitudinal coordinates of non-equal rotational parameters. Therefore, we will treat $y$ and $x_1\ldots x_p$ in the same way:
\be
y_a=(x_1,\ldots,x_p, y),\qquad \widetilde{m}_a=(m_1,\ldots,m_p, m_{p+1})\;:\quad \sum_{a=1}^{p+1}\frac{1}{\widetilde{m}_a}=1,\quad  \sum_{a=1}^{p}\frac{y_a^2}{\widetilde{m}_a}+{\frac{y^2}{\widetilde{m}_{p+1}}}=1,
\ee
in terms of which the metric takes the form
\be
\label{eq:metric_intermediate_special}
\frac{ds^2}{r^2_H}=A(y)\left(-r^2d\tau^2+\frac{dr^2}{r^2}\right)+
 dy_{p+1}^2+y_{p+1}^2d\Omega_{l-1}+\sum_{a=1}^{p}
(dy_a)^2 +
\sum_{i,j=1}^N\tilde{\gamma}_{ij}x_i(y) x_j(y) D\varphi^iD\varphi^j,
\ee
with
\be
A(y) ={\frac{\sum_{a=1}^{p+1} y^2_a/\widetilde{m}^2_a}{4\sum_{a<b}^{p+1}\frac{1}{\widetilde{m}_a}\frac{1}{\widetilde{m}_b}}}, \qquad \sum_{a=1}^{p+1}\frac{y_a^2}{\widetilde{m}_a}=1.
\label{metspec}\ee

Hamiltonian of the corresponding spherical mechanics then reads
	\be\label{eq:red_mech_inter_special}
	\mathcal{I}=A\left[\sum_{a,b=1}^{p}h^{ab}p_a p_b+\sum_{a=1}^{p+1}\frac{g_a^2}{y_a^2}+g_0\right],\quad {\rm with}\quad
	g_a^2=(p^2_{\varphi_1},\ldots, p^2_{\varphi_p}, \mathcal{I}_{p+1}),\quad  h^{ab}=\delta^{ab}-\frac{1}{\sum\limits_{a=1}^{p+1} y^2_a/\widetilde{m}^2_a}\frac{y_a}{\widetilde{m}_a}\frac{y_b}{\widetilde{m}_b}.
\ee
and $\mathcal{I}_{p+1}$ defined as  by \eqref{oddIsoHam} in $(p+1)$-dimensional space.
The above describes a  lower-dimensional version of \eqref{L}, where all rotational parameters are nonequal and we can analyze it as we did for the general case in the previous section. That is,
we introduce on the $(p+1)$-dimensional ellipsoidal coordinates
\be
	\label{eq:ellips_intremediate_special}
	y_a^2=\frac{\prod_{b=1}^{p+1}\left(\widetilde{m}_a-\lambda_b\right)}{\prod^{p+1}_{b=1; b\neq a}\left(\widetilde{m}_a-\widetilde{m}_b\right)},
\ee
and take $\lambda_{p+1}=0$ for  resolving the constraint \eqref{metspec} given by the second expression. The rest of the analysis goes through as  in \cite{non-equal-general}  and as in Section \ref{Particle-dynamics-Sec}.

The partially isotropic case discussed here, as we see, interpolates between the generic case of Section \ref{Particle-dynamics-Sec} ($p=N-1$) and the fully isotropic case ($p=0$) of section \ref{fully-isotropic-sec}: It decouples to the  Hamiltonians of type \eqref{eq:conf_ham_odd} and  \eqref{oddIsoHam}.
The case $l=1$ corresponds to  the system with non-equal parameters, and  the spherical subsystem is trivial ($\mathcal{I}_{p+1}=p_{\varphi_{p+1}}^2$).
For  $l\geq 2$ the $(l-1)$-dimensional  spherical subsystem is not trivial anymore and has $2(l-1)-1$  constants of motion.
Thus  the reduced $(N-1)$-dimensional angular system has $p+2l-3 =N-1+l-2$ constants of motion, i.e. the  number of extra constants of motion  compared to the generic case is $l-2$, with $l>2$.
It becomes maximally superintegrable  only for $l=N$, i.e when  all rotational parameters are equal.

This discussion can be easily extended to the case of even dimensions ($\sigma=1$). Here we will have an additional latitudinal coordinate ($p+l=N+1$) and a rotational parameter with a fixed value ($m_{N+1}=1$). One should note that $m_{N+1}$ cannot be equal to any other rotational parameter,  so it is one of the $p$ non-equal parameters.  In the limiting case when $l=1$ and all rotational parameters are different and we have an integrable system with $p=N$ configuration space degrees of freedom, as expected. Since $m_{N+1}$ cannot be equal to the others, $p$ cannot be equal to $0$ and the even dimensional system cannot be maximally superintegrable. In the limit when all rotational parameters are equal except $m_{N+1}$ ($p=1$), the system will lack one integral of motion to be maximally superintegrable.

\subsection{General case}
\label{sec:gc}

Having discussed the some equal $m_i$'s but the rest nonequal case, we now turn to the  most general case when there are  $s$ sets (blocks) of equal rotation parameters each containing $l_i$ members. As before we  assume that there are $p$ rotation parameters which are not equal to the others, so that $p+\sum_{i=1}^s l_i=N_\sigma$. Note that in our conventions $l_i\geq 2$. We introduce an upper index which, written on a parameter or a function, denotes the number of the block under consideration. So, for example $m_a^{(i)}$ will denote all the equal rotational parameters in the $i$-th set of rotation parameters and $x_a^{(i)}$ will denote their corresponding latitudinal coordinates and
\be
\begin{gathered}
	\{m_a^{(i)}\}=m_{p+l_1+\ldots +l_{i-1}+a}\equiv\kappa^{(i)}\qquad
	i=1,\ldots,s,\qquad a=1,\ldots,l_i.
\end{gathered}
\ee
The list of all rotational parameters can be written as
\be
\begin{gathered}
	\{m_\alpha\}=m_1,\ m_2,\ldots,\ m_p,\ \{m_a^{(1)}\},\ \{m_a^{(2)}\},\ldots,\ \{m_a^{(s)}\},\qquad \alpha=1,\ldots,N\\
	m_1\ne m_2\ne\ldots\ne m_p,\qquad \{m_a^{(i)}\}=\kappa^{(i)}\quad
	\text{with}\qquad \kappa^{(i)}\ne\kappa^{(j)}, \quad p+l_1+\ldots +l_{s}=N.	
\end{gathered}
\ee

Let us start with the odd  ($\sigma=0)$ case and the metric \eqref{m2}. We can construct the Hamiltonian for the reduced mechanics by introducing spherical and ellipsoidal coordinates. Different spherical coordinates will be introduced separately for each set of latitudinal coordinates corresponding to different sets of equal rotational parameters.
\be
\label{eq:shpere_intremediate}
x_1^{(i)}=r_i\prod_{\alpha=1}^{l_i-1}\sin{\theta_{\alpha}^{(i)}}\qquad	x_k^{(i)}=r_i\cos{\theta_{k-1}^{(i)}}\prod_{a=k}^{l_i-1}\sin{\theta_{a}^{(i)}}\qquad
x_{l_i}^{(i)}=r_i\cos{\theta_{l_i-1}^{(i)}},\qquad
k=2,\ldots, l_i-1
\ee
One should note that these spherical coordinates satisfy the  relations
\be
	\sum_{a=1}^{l_i}(x_a^{(i)})^2=r_i^2 \quad
	\text{and} \quad
	\sum_{a=1}^{l_i}(\diff x_a^{(i)})^2={\diff r_i}^2+r_i^2d \Omega^{(i)}_{l_i-1},
\ee
where $d\Omega^{(i)}_{n}=(\diff {\theta^{(i)}_{n}})^2+\sin^2\theta^{(i)}_{n}d\Omega^{(i)}_{n-1}$ denotes
 the metric on unit $n$-dimensional sphere.
For the rest of the latitudinal coordinates $x_1\ldots x_p$ corresponding to non-equal rotational parameters and the radial coordinates $r_i$ of  isotropic subsystems
 we introduce the notation
 \be
	\label{eq:ellips_intremediate}
	\{y_a\}=\{x_1,\ldots,x_p;\ r_1,\ldots, r_s\},\qquad
	\{\widetilde{m}_a\}=\{m_1,\ldots,m_p;\ \kappa^{(1)},\ldots, \kappa^{(s)}\},
\ee
In this notation the  the metric \eqref{m2} can be rewritten as
\be\label{eq:metric_intermediate}
	\frac{ds^2}{r^2_H}=A(y)\left(-r^2d\tau^2+\frac{dr^2}{r^2}\right)+\sum_{a=1}^{p+s} {dy_a}^2+\sum_{b=1}^sy_{p+b}^2d\Omega^{(b)}_{l_b-1}+
	\sum_{i,j=1}^N\tilde{\gamma}_{ij}x_i(y) x_j(y) D\varphi^iD\varphi^j,
\ee
where   $\tilde\gamma_{ij}$ , $A(y)$ are defined as in \eqref{gamma} and \eqref{metspec} respectively.
Therefore,  the Hamiltonian of the corresponding angular mechanics reads
\be
\begin{gathered}
	\label{eq:red_mech_inter}
	\mathcal{I}=A\left[\sum_{a,b=1}^{p+s-1}h^{ab}\pi_a\pi_b+\sum_{a=1}^{p+s}\frac{g_a^2}{y_a^2}+g_0\right]\qquad
	\{g_a^2\}=\{p^2_{\varphi_1},\ldots, p^2_{\varphi_p};\ \mathcal{I}^{(1)},\ldots,\mathcal{I}^{(s)}\}, \qquad \mathcal{I}^{(a)}=F^{(a)}_{l_a-1},	
\end{gathered}
\ee
where $\mathcal{I}^{(a)}$ are the spherical subsystems resulting from the $s$ sets of equal rotation parameters, $h^{ab}$ is defined by \eqref{eq:red_mech_inter_special}, and
\be
\begin{gathered}
	F_{d}^{(a)}=p_{\theta_{d}^{(a)}}^2+\frac{(g_{d+1}^{(a)})^2}{\cos^2\theta_{d}^{(a)}}+\frac{F_{d-1}^{(a)}}{\sin^2\theta_{d}^{(a)}}, \qquad
	F_{0}^{(a)}=(g_{1}^{(a)})^2, \qquad g_{d}^{(a)}=p_{\varphi_{p+l_1+\ldots+l_{a-1}+d}}\\
	\{\pi_a,\lambda_b\}=\delta_{ab},\qquad
	\{p_{\varphi_i},\varphi_j\}=\delta_{ij}\qquad
	\{p_{\theta_{b}^{(a)}},\theta_{d}^{(c)}\}=\delta_{ac}\delta_{bd},
\end{gathered}
\ee
Hence, the reduced spherical mechanics \eqref{eq:red_mech_inter} has the exact form of \eqref{eq:conf_ham_odd} (with appropriate constants) whose integrability has already been discussed.
 All discussions from the previous subsection can be easily extended to this case, e.g.  separation of variables
may be achieved in the ellipsoidal coordinates
\be
y_a^2=\frac{\prod_{b=1}^{{p+s}}\left(\widetilde{m}_a-\lambda_b\right)}{\prod^{{p+s}}_{b=1; b\neq a}\left(\widetilde{m}_a-\widetilde{m}_b\right)},
\ee
and place $\lambda_{p+s}=0$ for resolving the constraint on  latitudinal coordinates \eqref{eq:restr_on_x}, which  now takes the form
$\sum_{a=1}^{p+s}\frac{y_a^2}{\widetilde{m}_a}=1$.

So, we separated the variables for the $(N-1)$-dimensional angular mechanics describing the geodesics in the near-horizon limit of $(2N+1+\sigma)$-dimensional Myers-Perry black hole  in arbitrary dimension
with arbitrary non-zero values of rotational parameters.
The number of constants of motion in this system  can be easily counted: it is equal to $d+N_\sigma-p-2s$. The generic case of nonequal $m_i$ is recovered by $s=0, p=N_\sigma$ and the fully isotropic case as $s=1, p=0$.
In a similar manner one can construct associated Killing tensors.

\subsection{Contraction from fully non-isotropic to isotropic NHEMP  }\label{sec:contrraction}

Having the two corner cases discussed (fully non-isotropic and isotropic) an interesting question arises. What kind of approximation would transform the first integrals of fully non-isotropic NHMEP to the first integrals of isotropic NHEMP? It is straightforward to check that simply taking all rotation parameters to be equal just transforms all the first integrals of fully non-isotropic NHMEP to the Hamiltonian of the spherical mechanics of isotropic NHEMP (with an overall constant factor and a constant term). So if $m_i=N$
\be
	F_a=
	C_a\left(\sum\limits_{b,c=1}^{N-1}(\delta_{bc}-{x_b x_c})p_{b}p_{c}+\sum_{k=1}^N \frac{p^2_{\varphi_k}}{x_k^2}\right)+C^\prime_a
\ee
where $C_a$ and $C^\prime_a$ are constants. To find the desired approximation, we will work with rotation parameters which have little variations from their isotropic value $N$ ($\epsilon_i\ll N$), \[m_i=N+\epsilon_i.\] In such a limit, the Hamiltonian of the non-isotropic mechanics can be extended in powers of $\epsilon_i$, keeping the first order term only
\be
\begin{aligned}
	\label{eq:F_1}
	F_1=N^{N-3}\left[N\casimir_{iso}+N^2g_0
	-\sum_{i=1}^{N}\epsilon_ix_i^2 \left[\sum_{a}^{N-1}p_a^2+\sum_{k}^{N}\frac{p_{\varphi_k}^2}{x_k^2}+g_0\right]+2\sum_{a,b}^{N-1}\epsilon_a p_ax_ap_bx_b\right]
\end{aligned}
\ee
where
\be
	\casimir_{iso}=\sum\limits_{a,b=1}^{N-1}(N\delta_{ab}-{x_a x_b})p_{a}p_{b}+N\sum\limits_{i=1}^{N}\frac{p_{\varphi_i}^2}{x_i^2}
\ee
is the isotropic Hamiltonian. We should note that the linear term of $F_1$ still corresponds with the isotropic Hamiltonian $\casimir_{iso}$ but the relation $\sum x_i^2=N$ doesn't hold anymore.

Now, if we find some linear combination $P(F_a)$ of first integrals of non-isotropic mechanics such that the free term of the expansion around $m_i=N$ vanishes, we can write
\begin{equation}
\label{eq:iso_integ}
\begin{aligned}
\{P(F_a),F_1\}&=0=\left\{\sum_{i=1}^{N}\epsilon_iP_i(p_j,x_j)\ ,\ \casimir_{iso}+\sum_{i=1}^{N}\epsilon_i(...)\right\}=\sum_{i=1}^{N}\epsilon_i\left\{P_i(p_j,x_j)\ ,\ \casimir_{iso}\right\}\\
&\implies \qquad  \left\{P_i(p_j,x_j)\ ,\ \casimir_{iso}\right\}=0
\end{aligned}
\end{equation}
We see  that the first order coefficients $P_i(p_j,x_j)$ of the $P(F_a)$ linear combination are first integrals for $\casimir_{iso}$. To construct such combination whose free term vanishes we can take any of the first integrals, let's say $F_{N-1}$ and expand it.
\be
\label{eq:F_N-1}
\begin{aligned}
	F_{N-1}=(-1)^N&\left[\casimir_{iso}-\frac{g_0}{N}\sum_{i=1}^{N}\epsilon_ix_i^2+\sum_{i}^{N}\epsilon_i\frac{p_{\varphi_i}^2}{x_i^2}+\sum_{a=1}^{N-1}\epsilon_a p_a^2\right]
\end{aligned}
\ee
We see from \eqref{eq:F_1} and \eqref{eq:F_N-1} that by combining $F_1$ and $F_{N-1}$ the free term can be eliminated
\be
\label{eq:P_comb}
\begin{aligned}
	N^{-(N-3)}F_1&+(-1)^{N-1}NF_{N-1}-g_0N^2=\\
	&-\left(\sum_{a}^{N-1}p_a^2+\sum_{k}^{N}\frac{p_{\varphi_k}^2}{x_k^2}\right)\sum_{i=1}^{N}\epsilon_ix_i^2+2\sum_{a,b}^{N-1}\epsilon_a p_a x_a p_b x_b -N\left(\sum_{i}^{N}\epsilon_i\frac{p_{\varphi_i}^2}{x_i^2}+\sum_{a=1}^{N-1}\epsilon_a p_a^2\right)
\end{aligned}
\ee
Furthermore, from the expression $\sum_{i}^{N}x_i^2/m_i=1$ we can find \[x_N^2=\left(\tilde{x}_N^2+\frac{1}{N}\sum_{a}^{N-1}\epsilon_ax_a^2\right)\left(1+\frac{\epsilon_N}{N}\right),\quad \tilde{x}_N^2\equiv N-\sum_{a}^{N-1}x_a^2\]
and replace with this relation every occurrence  of $x_N$ in \eqref{eq:P_comb}. Doing this, we will end up with the same equation \eqref{eq:P_comb} with just $x_N^2$ replaced by $\tilde{x}_N^2$. So in further calculations we are free to consider equation \eqref{eq:P_comb} with a redefined $x_N$
\be
\tilde{x}_N^2 \rightarrow x_N^2=N-\sum_{a}^{N-1}x_a^2
\ee
Thus, having in mind \eqref{eq:iso_integ}, we find the first integrals of isotropic mechanics to be
\be
\begin{aligned}
	F_a^{iso}&=-x_a^2\left(\sum_{b}^{N-1}p_b^2+\sum_{k}^{N}\frac{p_{\varphi_k}^2}{x_k^2}\right)+2 p_a x_a \sum_{b}^{N-1}p_b x_b -N\left(\frac{p_{\varphi_a}^2}{x_a^2}+ p_a^2\right)\\
	F_{N}^{iso}&=-x_N^2\left(\sum_{b}^{N-1}p_b^2+\sum_{k}^{N}\frac{p_{\varphi_k}^2}{x_k^2}\right)-N\frac{p_{\varphi_N}^2}{x_N^2}
\end{aligned}
\ee
Now, we can see that the sum of all $N$ first integrals results into the casimir of isotropic mechanics
\be
\sum_{i=1}^NF_i^{iso}=-2\ \casimir_{iso}.
\ee
Thus, by definition, all $F_i^{iso}$ commute with $\sum_{i=1}^NF_i^{iso}$, but one can check that they don't commute with each other.

\section{Extremal vanishing horizon case}\label{EVH-Sec}
As  seen from metric \eqref{NHEMP-metric-rt-coord}, the case where one of the $a_i$'s is zero is a singular case. In fact for this case one should revisit the near-horizon limit. It has been shown that \cite{NHEVH-MP} for the odd dimensional extremal MP black holes the horizon area also vanishes and we are hence dealing with an Extremal Vanishing Horizon (EVH) black hole \cite{NHEVH-1}. The near-horizon EVH black holes have remarkable features which are not shared by generic extremal black holes; they constitute different set of geometries which should be studied separately \cite{NHEVH-three-theorems}. In particular, it has been proved that for EVH black holes the near horizon geometry include an AdS$_3$ factor (in contrast with the AdS$_2$ factor of general extremal case) \cite{NHEVH-three-theorems, Sadeghian:2017bpr}, i.e. the $d$ dimensional NHEVHMP exhibits $SO(2,2)\times U(1)^{N-1}$ isometry.
 To study this case, we start by a review on black hole geometry itself.
  Then, by taking the near horizon and EVH limit, we discuss the separability of
  Hamilton-Jacobi equations on the NHEVH geometries.

As discussed in the special case of EVH black holes one has to revisit the standard NH theorems for extremal black holes. Here we review EVH black holes in the  family of odd dimensional MP black holes \cite{mp}:
\be\label{MP-odd}
	ds^2=-d\tau^2+\frac{\mu \rho^2}{\Pi F}(d\tau+\sum_{i=1}^N a_i \mu_i^2d \phi_i)^2+\frac{\Pi F}{\Pi-\mu \rho^2} d\rho^2+\sum_{i=1}^N (\rho^2+a_i^2)(d\mu_i^2+\mu_i^2d\phi_i^2)
\ee
where
\be
F=1-\sum_i\frac{a_i^2\mu_i^2}{\rho^2+a_i^2},\quad \Pi=\prod_{i=1}^N (\rho^2+a_i^2),\qquad \sum_i \mu_i^2=1.
\ee
The extremal case happens when $\Pi-\mu \rho^2=0$ has double roots and the EVH case is when one of $a_i$ parameters, which we take to be $a_N$ is zero. That is in the EVH case $\mu=\prod_{a=1}^{N-1} a_a^2$. We note that we could have considered  a ``near-EVH'' metric where the black hole is at a non-zero but small temperature and the horizon area is also small, while the ratio of horizon area to the temperature is finite \cite{NHEVH-1, NHEVH-three-theorems}.

The horizon for the EVH case is at $\rho=0$ and hence in the NH limit, the leading contributions come from
\be\label{NHEVH-limit}
	\Pi=\mu \rho^2 (1+\frac{\rho^2}{r_0^2}),\qquad F_0=1-\sum_{a=1}^{N-1} \mu_a^2,\qquad \frac{1}{r^2_0}=\sum_{b=1}^{N-1}\frac{1}{a_b^2}.
\ee
Plugging the above into the metric \eqref{MP-odd} and taking:
$$\rho=r_0\, r\,\epsilon,\quad \tau=r_0\, t/\epsilon,\quad \psi=\varphi_N/\epsilon,\quad \varphi_a=\phi_a+\tau/a_a,\quad a=1,\ldots, N-1, \qquad \epsilon\to 0,$$
we obtain the NHEVHMP metric \cite{NHEVH-MP}:
\be\label{NHMPEVH-metric}
	ds^2=F_0\,r^2_0\,\left[-r^2\, dt^2+\frac{dr^2}{r^2}+r^2d\psi^2\right]+ \sum_{b =1}^{N-1}a_b^2d\mu_b^2+\sum_{a,b =1}^{N-1}\gamma_{ab}d\varphi_ad\varphi_a,\qquad\gamma_{ab}\equiv a^2_a \mu_a^2\delta_{ab}+ a_a a_b \frac{\mu_a^2\mu_b^2}{F_0}.
\ee
where in the above $a,b$ run from $1$ to $N-1$.
Had we started from the near-EVH geometry, the AdS$_3$ factor (the $r,t,\psi$ part) of \eqref{NHMPEVH-metric} would have turned into a generic BTZ black hole geometry \cite{NHEVH-1, NHEVH-three-theorems}. The NH geometry \eqref{NHMPEVH-metric} has $SO(2,2)\times U(1)^{N-1}\simeq SL(2,\mathbb{R})\times SL(2,\mathbb{R})\times U(1)^{N-1}$ isometry. This is to be compared with $SL(2,\mathbb{R})\times U(1)^N$ of the non-EVH NHEMP discussed in previous sections.

To discuss separability of the particle dynamics on \eqref{NHMPEVH-metric}, as in the previous sections, we introduce coordinates,
\be
	 x_a\equiv \frac{a_a \mu_a}{r_0}\,\quad m_a\equiv \frac{a_a^2}{r^2_0}\, \qquad \sum_{a=1}^{N-1}\frac{1}{m_a}=1,
\ee
in which  \eqref{NHMPEVH-metric} takes the form
 \be\label{rEVH}
	 \frac{ds^2}{r^2_0}=F_0ds^2_{AdS_3}+\sum_{a}^{N-1} dx_a^2 +\sum_{a,b}^{N-1} \tilde{\gamma}_{ab}x_ax_bd\varphi_ad\varphi_b,
 \ee
 with
 \be
 \begin{gathered}
	 ds^2_{AdS_3}=
	 r^2\,\left(-dt^2+{d\psi^2}\right)+\frac{ dr^2}{r^2},
	 \qquad  F_0=1-\sum_a^{N-1}\frac{x^2_a}{m_a},\\
	 \tilde{\gamma}_{ab}x_ax_b=\frac{1}{r_0^2}\gamma_{ab},
	 \qquad \tilde{\gamma}_{ab}=\delta_{ab}+\frac{1}{F_0}\frac{x_a}{\sqrt{m_a}}\frac{x_b}{\sqrt{m_b}}.
 \end{gathered}
 \ee
%
%
%
%
%
%
%
%
%
%
The generators of the two $SL(2,\mathbb{R})$ Killing vectors may be written as
\bea
	&&H_+={\partial_v}\,,\qquad D_+=v\,{\partial_v}-r\,\partial_r\,\qquad K_+=v^2\,{\partial_v}+\frac{1}{r^2}\,{\partial_u}-2r\, v\,{\partial_r}\,,\nnr
	&&H_-={\partial_u}\,,\qquad D_-=u\,{\partial_u}-r\,\partial_r\,\qquad K_-=u^2\,{\partial_u}+\frac{1}{r^2}\,{\partial_v}-2r\, u\,{\partial_r}\,,
\eea
where $v=t+\psi$ and $u=t-\psi$.  The Casimir of $SL(2,\mathbb{R})$'s are
 \bea
\mathcal{I_\pm}=H_\pm \,K_\pm-D_\pm^2
\eea
and one can readily check that both Casimirs are equal to $\mathcal{I}=\frac{1}{r^2}\left(\partial_t^2-\partial_{\psi}^2\right)-r^2\,\partial_{r}^2\,.$

The mass-shell equation of the probe particle \eqref{mass-shell-1} then reads
\be
\label{kg}
	\frac{(p_0)^2 -(p_\psi)^2}{r^2}=(rp_r)^2 +\mathcal{I}(p_a,x_a, p_{\varphi_a})
\ee
where
\be
\{p_a, x_b\}= \delta_{ab},\quad \{p_{\varphi_a},\varphi^b\}=\delta_{ab},\quad \{p_\psi,\psi\}=1, \quad\{p_r, r\}=1,
\ee
and
\be
	\label{nullI}
	\mathcal{I}(p_a,x_a, p_{\varphi_a})=
	(1-\sum_{c=1}^{N-1}\frac{x^2_c}{m_c})
	\left[\sum_{a=1}^{N-1}{p^2_a}+\sum_{a=1}^{N-1}\frac{p^2_{\varphi_a}}{x_a^2}+g_0\right],
	\qquad g_0=-\left(\sum_a^{N-1}\frac{p_{\varphi_a}}{\sqrt{m_a}}\right)^2+m_0^2r_0^2,
\ee
where  ${\cal I}$ in \eqref{nullI} is the Casimir. Note that while the background has  $SL(2,\mathbb{R})\times SL(2,\mathbb{R})\times U(1)^{N-1}$ isometry the Casimirs of the the two $SL(2,\mathbb{R})$ factors happen to be identically the same and hence we are dealing with a single ${\cal I}$; appearance of an extra $SL(2,\mathbb{R})$ does not add to number of constant of motion compared to the non-EVH case.

Hence, as in the regular case, we have to consider separately three cases

\begin{itemize}

\item {\sl Generic, non-isotropic case, all $m_a$ are non-equal}

To separate the variables in \eqref{nullI}, in the special case when none of the rotational parameter is equal, we introduce the ellipsoidal coordinates
\bea
	x_a^2=\frac{\prod_{b=1}^{N-1}(m_a-\lambda_b)}{{\prod_{b\ne a}^{N-1}}(m_a-m_b)}.
\eea
In this terms the angular Hamiltonian reads
\be
	\label{eq:vanishing_none_eq}
	\mathcal{I}=\left(\prod_a^{N-1}\frac{\lambda_a}{m_a}\right)
	\left[\sum_{a=1}^{N-1}\frac{4\prod_{b}^{N-1}(m_b-\lambda_a)}{\prod_{b\ne a}^{N-1}(\lambda_b-\lambda_a)}\pi_a^2+\sum_a^{N-1}\frac{p_{\varphi_a}^2}{x_a^2}+g_0\right],
\ee
where $\{\pi_a,\lambda_b\}=\delta_{ab}$. One can see that \eqref{eq:vanishing_none_eq} has a very similar form to  \eqref{eq:conf_ham_odd}, and using the identities  \eqref{26} and \eqref{eq:relation_1}, it can be rewritten as follows (after fixing the Hamiltonian $\mathcal{I}=\mathcal{E}$)
\be	
	\label{eq:vanishing_R}
	\sum_{a=1}^{N-1}\frac{R_a- \mathcal{\tilde{E}}}{\lambda_a\prod_{b=1,a\ne b}^{N-1}(\lambda_b-\lambda_a)}=0,
\ee
where
\be	
\begin{gathered}
	R_a=4\lambda_a\pi_a^2\prod_{b}^{N-1}(m_b-\lambda_a)+(-1)^{N-1}\sum_b^{N-1}\frac{\lambda_a g_b^2}{\lambda_a-m_b}-g_0(-\lambda_a)^{N-1},\\
	g_a^2=p_{\varphi_a}^2\prod_{b=1}^{N-1}(m_a-m_b),\qquad
	\mathcal{\tilde{E}}=\mathcal{E}\prod_a^{N-1}m_a.
\end{gathered}
\ee
Separation of variables and the constants of motion is similar to the section \ref{Particle-dynamics-Sec}, where \eqref{eq:vanishing_R} corresponds to \eqref{HJO}.

\item {\sl Isotropic case, all $m_a$ are equal}

In this case ($m_a=N-1$), we separate the variables in \eqref{nullI} by introducing spherical coordinates $\{u,\ ,y_\alpha,\ \theta_{N-2}\}$
\be
	x_{N-1}=u \cos \theta_{N-2}, \qquad x_{N-1-\alpha}=u\ y_\alpha \sin \theta_{N-2},\qquad
	\sum_{\alpha=1}^{N-2}y_\alpha ^2=1
\ee
where $\alpha=1\ldots N-2$. In these coordinates \eqref{nullI} will take the following form
\be
	\mathcal{I}=\left(1-\frac{1}{N-1}u^2\right)\left[p_u^2+\frac{F_{N-2}}{u^2}+g_0\right]
\ee
with $F_a$ defined in \eqref{eq:iso_integrals_odd}, where the separation of variables and the derivation of integrals of motion was carried out according to \ref{fully-isotropic-sec}.


\item {\sl partially isotropic case}

The last case is the most general one which involves  sets of equal  and a set of non-equal rotational parameters. With the discussions of the two previous cases (fully isotropic and fully non-isotropic) in view and  recalling the analysis of partially isotropic NHEMP case of previous section, it is  straightforward to separate the variables in partially isotropic NHEVHMP. Following the steps in section \ref{sec:gc}, one should first introduce different spherical coordinates for each set of equal rotational parameters and ellipsoidal coordinates for the joint set of non-equal rotational parameters and the radial parts of spherical coordinates. This will result into a spherical mechanics similar to \eqref{eq:red_mech_inter} where the Hamiltonians of spherical subsystems will be included as parameters.

\end{itemize}

\section{Discussion}
In this work, continuing analysis of \cite{non-equal-general, Demirchian:2017uvo}, we studied  separability of geodesic motion on the near horizon geometries of Myers-Perry black hole in $d$, even or odd, dimensions and established the integrability by explicit construction of $d$ constants of motion.  In the general case $[\frac{d-1}{2}]+1$ of these constants of motion are related to the Killing vectors of the background (note that the background in general has $[\frac{d-1}{2}]+3$ Killing vectors, but three of them form an $sl(2,\mathbb{R})$ algebra and hence there is only one independent  conserved charge from this sector). Our analysis reconfirms the earlier observations that although near-horizon limit in the extremal black holes enhances the number of Killing vectors by two \cite{NHEG-general}, the number of independent conserved charges from the Killing vectors does not change.
Our system, in the general case,  has $[\frac{d}{2}]$ constants of motion are associated with second rank Killing tensors the system possesses. We also constructed  the explicit relation between these Killing tensors  and the conserved charges and one may check that our Killing tensors and those in  \cite{Hidden-symmetry-NHEK} match. We note that the Killing tensors of \cite{Hidden-symmetry-NHEK} were obtained using the near horizon limit  on the Killing tensors of Myers-Perry black hole in a coordinate system which makes the geodesics of black hole separable itself. Whereas, we directly worked with ellipsoidal coordinates for the NHEMP,  introduced in \cite{non-equal-general}. Comparing the two systems before and after the NH limit,
 it was argued in  \cite{Hidden-symmetry-NHEK} that a combination of Killing tensors is reducible to the Killing vectors, however, we obtain other second rank Killing tensors, through which the system remains integrable.
Moreover, by explicitly showing the separability, one concludes that  there is no inconsistency with the theorems in \cite{Benenti and Francaviglia}. There is an extra conserved charge related to the Casimir of $SL(2,\mathbb{R})$ symmetry group which intrinsically exists in the NHEG's. We have shown that the charge of the Casimir is independent of the other conserved charges. In this sense, one of the ``hidden symmetries,'' symmetries which are associated with equations of motion and are not isometries of the background,  becomes explicit in the NH limit \cite{Hidden-symmetry-NHEK}.


Following the discussions in \cite{GNS-1}, we showed that for special cases where some of the rotations parameters of the background are equal, the  geodesic problem on NHEMP is superintegrable. We established superintegrability by establishing existence of  other constants of motion.  Our methods here, combined with those in \cite{GNS-1}, allows one to read the extra second rank Killing tensors obtained in these cases. The rough picture is as follows: We started with a system with $2N+1+\sigma$ variables with $N$ isometries. Fixing the momenta associated with the isometries, we obtained and focused the $N-1+\sigma$ dimensional ``angular mechanics'' part.  In this sector, whenever $N$ number of rotation parameters $m_i$ of the background metric are equal the $U(1)^N$ isometry is enhanced to $U(N)$ and this latter brings about other second rank Killing tensors. All in all, the fully isotropic case in odd dimensions with $U(\frac{d-1}{2})$ isometry, the $d-2$ dimensional spherical mechanics part  is maximally superintegrable, it has $N+(N-2)=2N-2$ extra constants of motion. The fully isotropic case in even dimensions, however, is not maximally superintegrable; it has still $2N-1$  extra Killing tensors (one less than the $N$ constants of motion to make the system fully superintegrable).  We discussed the ``special cases'' in two different ways. First, we reanalyzed the system from the scratch (in section \ref{fully-isotropic-sec}) and also took the equal rotation parameter limit of the generic case (in section \ref{sec:contrraction}). As expected, these two cases matched. Our preliminary analysis, which we did not show here, indicate that the above statements is also true for the NH limit of extremal MP black holes in (A)dS backgrounds.

We also discussed the EVH case, which happens for odd dimensional extremal MP when one of the  rotation parameters $a_i$ vanishes. In the general NHEVHMP case, where the background isometry is $SO(2,2)\times U(1)^{\frac{d-3}{2}}$ the number of independent charges associated with Killing vectors is $\frac{d+1}{2}$. Despited enhancement of the isometry group compared to the generic NHEMP case), we found that this symmetry enhancement does not add to number of independent constants of motion, the system in general does not pose extra constants of motion and remains just integrable.





Here we explored second rank Killing tensors, one may suspect is the system has independent higher rank Killing tensors too. Although it is unlikely, if it happens the system for the generic rotation parameters becomes superintegrable. It is interesting to explore this question.
Finally, as already pointed out in the introduction, one can consider other probes including scalar, Dirac field or gauge or tensor perturbations on the NHEMP backgrounds and study their integrablitiy. To this end, the study of Killing Yano tensor and principal tensor \cite{Frolov:2007nt,Kubiznak:2006kt} should be completed. We hope to address this in our future publications.

\acknowledgments
 The work of S.S. and M.M.Sh-J. is supported in part
by  the junior research chair in black hole physics of Iranian NSF, grant number 950124. A.N. and H.D. would like to thank School of Physics (IPM, Tehran) for hospitality during the course of this project.
All the authors especially and gratefully acknowledge  support of ICTP program network scheme NT-04. H.D. acknowledges  The ICTP   Affiliated Center program AF-04 and
Volkswagenstiftung. This work was made possible in part by a research grant from the Armenian National Science and Education Fund (ANSEF) based in New York, USA.

\end{document}